\documentstyle[aps,multicol]{revtex}
\begin{document}
\draft
\title{Short-time scaling in the critical dynamics of an antiferromagnetic 
Ising system with conserved magnetisation}
\author{Parongama Sen and Subinay Dasgupta}

\address{Department of Physics, University of Calcutta,
92 A.P. C. Road, Calcutta 700009, India.\\
e-mail : paro@cubmb.ernet.in, subinay@cubmb.ernet.in}
\maketitle
\begin{abstract}
We study by Monte Carlo simulations the short-time exponent $\theta$ in an
antiferromagnetic Ising system for which the magnetisation is conserved but
the sublattice magnetisation (which is the order parameter in this case) 
is not.
This system belongs to the dynamic class of model C. We use nearest neighbour
Kawasaki dynamics so that the magnetisation is conserved {\em locally}.
We find that in three dimensions $\theta$ is independent of the conserved
magnetisation. This is in agreement with the available theoretical studies, but in
disagreement with previous simulation studies with global conservation
algorithm. However, we agree with both these studies regarding the result
$\theta_C \ne \theta_A$. We also find that in two dimensions, 
$\theta_C = \theta_A$.
\end{abstract}

\begin{multicols}{2}
\bigskip

In equilibrium statistical physics, universal scaling laws are 
observed close to a critical point where the correlation length diverges. 
Dynamical systems also exhibit a universal scaling behaviour in the long time 
regime. Dynamic universility classes are  characterised by the dynamic 
exponent, which connects the divergences in space and time.  
Typically, in a magnetic system,  the finite size scaling form of a physical observable 
$O(t,\tau,L)$ is given by,
\begin{equation}
O(t,\tau,L) = b^{-x}O(b^{-z}t,b^{1/\nu}\tau,b^{-1}L)
\end{equation}
where $\tau
= (1-T/T_c)$ 
 is the deviation from the critical temperature 
$T_c$, 
$b$ the scaling factor, 
$\nu$ and $x$ the static critical exponents, 
$z$ the dynamic exponent and $L$ is the linear size of the system.

Some time ago, it was found that there exists
a universal behaviour in short-time regime as well\cite{jss}.  
If a  magnetic system is quenched from a high temperature
to its critical temperature $T_c$, with initial order parameter
equal to $m_0$, 
then  universal scaling behaviour is observed in a macroscopic 
short-time regime :

\begin{equation}
M(t,\tau,L,m_0) = b^{-\beta/\nu} M(b^{-z}t,b^{1/\nu}\tau,b^{-1}L,b^{x_0}m_0)
\end{equation}
where $M$ is the order parameter, $t$ is time, $\beta$ is the critical exponent 
associated with the order parameter
and $x_0$ is a new exponent associated with the short-time 
effect.
$x_0$ is the scaling dimension of $m_0$. 
At the critical point,  for {\it small} values of $m_0$,
\begin{equation}
M(t,m_0) \sim m_0t^\theta
\end{equation}
with $\theta = (x_0 -\beta/\nu)/z$.
$\theta $ is a new exponent, not 
related to any previously known static or dynamic exponent \cite{jss}. The phenomenon is universal
as $\theta$ does not depend on microscopic details,
updating schemes etc. It is observed in a macroscopic short
time regime.

A positive value of $\theta$ will  indicate that the magnetisation 
(order parameter in general)
will first increase in time and later show the conventional
power law decrease.  It was observed in many systems that such an
increase indeed occurs. The exact cause of this behaviour is not yet
known very clearly.

 Systems with different critical dynamics have been classified as 
models A, B, C etc \cite{HH}. Model A has no conservation, in model B the order parameter is conserved
and in model C, the non-conserved order parameter  is coupled to a 
non-ordering conserved field. 
These dynamic classes are distinguished by the different values of the dynamic
exponent $z$.  

Using field theoretical methods, it was shown in \cite{jss} that short 
time behaviour exists in Model A.
Extensive numerical studies for calculating $\theta$ in the different 
dynamical classes have also been made in the recent years and
accurate estimates of $\theta$ in model A  
are available in different dimensions \cite{zhengrev}.
The results confirm the qualitative
behaviour predicted by the theoretical analysis. 
Systems belonging to the class of model B, 
do not show any short-time effect. However, numerical studies 
with a globally conserved order parameter in two dimensions indicate that
there could be a short time effect \cite{zheng1}. 

In model C, field theoretic techniques have again shown
that there is a universal short-time behaviour \cite{OJ}. 
Short time effect in a semi infinite system belonging to model C has also
been studied \cite{sutapa}.
There has  been some recent numerical studies of the short-time scaling 
in model C with global conservation \cite{zheng2}.
These numerical studies show the novel result that the short-time 
exponent depends on the value of the globally conserved 
magnetisation - a feature not obtained at least in the theoretical
study with local conservation. However, this is not very surprising,
as systems with global conservation in general show different 
behaviour compared to the ones with local conservation \cite{bray,satya,PS} as
far as the long time behaviour is concerned.
In model C and model B for example, the behaviour with  global conservation 
becomes model A-like. 

However, before any conclusive statement about the 
difference in short-time behaviour in model C with local and
non-local conservation can be made, it is necessary to obtain the 
numerical estimates of the exponents 
where the coupling field is locally conserved.
This is because the numerical estimates and field theoretic 
results could be quite different below the upper critical dimension.
In this communication, we report the estimates of the 
short-time exponents in both two and three dimensions 
with several values of the conserved density from simulation studies with
local conservation.



We have taken an antiferromagnetic Ising system where the magnetisation $m_0$ is kept
constant. This model has been studied numerically previously to estimate 
the dynamic 
exponent $z$ \cite{SDS} and the short-time exponent $\theta$ 
with global conservation 
\cite{zheng2}. The order parameter in this system is the 
staggered magnetisation $m_s$ 
which is not conserved. Initially the system is allowed to have short range 
correlations
only (since we quench it from a high temperature to $T_c$) and the initial $m_s(t=0)$ is kept at a low value. 
Hence the system is prepared with two constraints: both
$m_0$ and $m_s$ are kept fixed. This could be done by 
keeping a staggered field, but we simply keep it at the desired value by
appropriately flipping spins in a random configuration (the method is
analogous to what is called a sharply prepared state in \cite{zhengrev}).
The antiferromagnetic interactions are between nearest neighbours. The
lattices are hypercubic. Periodic boundary condition along
one direction and helical boundary condition along the other directions are
used. Helical boundary condition with antiferromagnetic nearest neighbour
intercations demand that the linear size along  $(d-1)$  
directions should be odd where $d$ is the spatial dimension. 

We use the Kawasaki dynamics where the opposite spins on 
nearest-neighbouring sites are
exchanged, thus keeping $m_0$ a constant. The exchange between
local neighbours ensure that the conservation is local.
The spins are updated sequentially and a sweep through the entire system 
is equivalent to one unit of time (one Monte Carlo step).
In both two and three dimensions, we fix  $m_0$ at several different values.
Since the system is quenched to the critical temperature $T_c(m_0)$, we first 
estimate the value of $T_c(m_0)$ (Table 1) for a large system following  the  
method used 
in \cite{SDS}. 
We then estimate $\theta$ for each  value of $m_0$ in both two and three 
dimensions in smaller lattices. For this estimate, one should measure the 
initial
slope of $m_s(t)$ vs $t$ curve for different values of $m_s(0)$ and take
the limit $m_s(0) \rightarrow 0$.


The results are as follows. In {\it two dimensions}, the short-time 
exponent $\theta$ has been found to be 
\begin{equation}
\theta = 0.18 \pm 0.01
\end{equation}
and this value has no measurable dependence on the initial (small) value of
sublattice magnetisation $m_s(0)$ (Fig. 1) and on the conserved 
magnetisation $m_0$ (Fig. 2). Also, in contrast with the simulation studies 
with global conservation algorithm \cite{zheng2}, 
it is the same for model A and C, upto 
the accuracy of the present study. In {\it three dimensions}, the 
exponent $\theta$ has been found to be
\begin{equation}
\theta _C = 0.13 \pm 0.01
\end{equation}
and this value also has no measurable dependence on the initial (small) value of
sublattice magnetisation $m_s(0)$ (Fig. 3). In agreement with the theoretical
predictions \cite{OJ}  and in contrast with the simulation studies with global 
conservation algorithm \cite{zheng2}, this value of $\theta$ has no numerically detectable
dependence on the conserved
magnetisation $m_0$ (Fig. 4) at least for $m_0 \ge 0.12$. 
However, in agreement with both the studies
\cite{OJ,zheng2}, the exponent has a different value for model A :
\begin{equation}
\theta_A = 0.10 \pm 0.01
\end{equation}
There is some indication that for very small values of $m_0$ ($\le 0.08$),
$\theta$ might depend on $m_0$ (Fig. 4) but  a detailed study in this
region requires very large scale simulations.
 
A few comments are in order.\\ (i) Our estimates of $\theta$ in 2$d$ 
for model A is in agreement with previous estimates \cite{zhengrev}. 
The estimates for model A are obtained by putting $m_0 = 0$ in the
present model (as $m_0 = 0$ corresponds to model  A \cite{eisen}). \\
(ii) That $\theta_A=\theta_C$ in 2$d$ indicates that the conservation is 
irrelevant here. This is because the specific heat exponent $\alpha$ is
negative here \cite{HH,OJ}, and the estimate of $\theta_C$ 
given by \cite{OJ} is strictly true for $2 < d <4$
when the spin dimension  $n=1$. \\
(iii) In 3$d$, our results 
$\theta_A \ne \theta_C$ and independence of $\theta_C$ of $m_0$ are in 
qualitative agreement with the theoretical estimates \cite{OJ}. \\
(iv) As regards the dependence of $\theta_C$ on $m_0$ in 3$d$, the 
discrepancy between our results and that
of \cite{zheng2} is not surprising, because non-local conservation may be
expected to give new values of exponents and can even
change the universality classes. 
In fact, the numerical estimate of the dynamical exponent $z$ in the
globally conserved model C \cite{zheng2} is found to
correspond to that of model A, a result predicted by the analytical study
of \cite{PS}. \\
(v) The results shown in the figures are for the largest sizes 
simulated. Simulation for smaller sizes, e.g., $199\times 200$ in 2$d$ and $41\times 41 
\times 42$ in 3$d$  
 show that there is no detectable finite size effect. \\
(vi) There could be some error in the estimate of $\theta$ due to
the error in the estimation of $T_c$. We have, however, checked that
$\theta$ does not vary measurably when $T_c$ is varied about it's
estimated value within the error bar.

Lastly, we should mention that one can also  estimate $\theta$
from the behavior of the autocorrelation function
\begin{equation}
A(t,0) = \langle S_i(t)S_i(0)\rangle 
- \langle S_i(t)\rangle \langle S_i(0)\rangle 
\sim t^{-\lambda}
\end{equation}
where $\lambda = d/z - \theta$. In this method, one can put
$m_s(0) =0$. 
One needs to know $z$ very accurately to have an
accurate estimate of $\theta$ using this method. Also, a good estimate of 
$\theta$  
requires  lattice sizes much greater than the ones simulated in the present study.
Nevertheless, we could  calculate  $A(t,0)$ and  verify that 
 $\lambda$ (and hence $\theta$) 
is indeed independent
of $m_0$.


The authors are grateful 
to S. M. 
Bhattacharjee for valuable discussions and suggestions and 
to B. Zheng for useful comments.
The computational facilities were provided by CUCC. PS acknowledges
financial support from DST project SP/S2/M-11/99.

 \newpage

\begin{center}

                   {\bf  Table 1 } \\
Numerical estimate of critical temperature $T_c$ for different values of 
the conserved magnetisation~$m_0$. The estimates for 3d with $m_0 \ne 0$
are in agreement with \cite{zheng2}. \\

\begin {tabular}{|c|c|c|c|}  \hline \hline
Dimension & Size of Lattice & $m_0$ & $k_B T_c$ \\ \hline \hline
2 & $1001\times1002$ & 0   & $2.2692^a$ \\ \cline{3-4}
  &                  & 0.1 & 2.108 $\pm 0.002$  \\ \cline{3-4}
  &                  & 0.2 & 1.557 $\pm 0.002$  \\ \hline
3 & $101\times101\times102$ & 0   & $4.5116^b$ \\ \cline{3-4}
  &                         & 0.08 & 4.453 $\pm 0.004$  \\ \cline{3-4}
  &                         & 0.12 & 4.376 $\pm 0.004$  \\ \cline{3-4}
  &                         & 0.2  & 4.142 $\pm 0.004$  \\ \cline{3-4}
  &                         & 0.4  & 3.009 $\pm 0.004$  \\ \hline \hline
\end{tabular}

$^a$ Exact Result. \hspace{.5cm}  $^b$ Approximate value, see e.g., \cite{zheng2}
\end{center}



\end{multicols}
 \pagebreak

\setlength{\unitlength}{0.240900pt}
\ifx\plotpoint\undefined\newsavebox{\plotpoint}\fi
\sbox{\plotpoint}{\rule[-0.200pt]{0.400pt}{0.400pt}}%
%
 \\
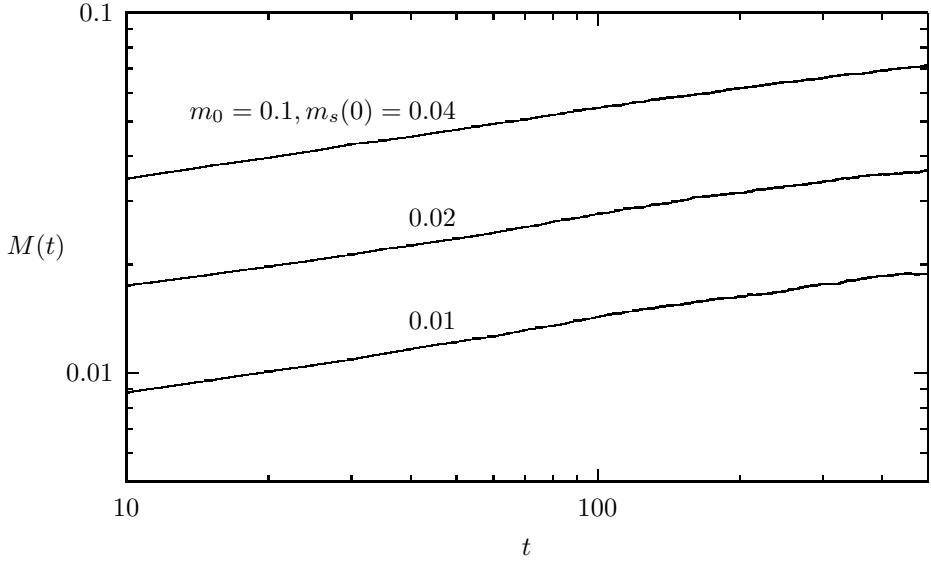
Fig. 1 : Early-time effect at critical temperature in two dimensions for a 
fixed value of the conserved
magnetisation $m_0$, and various values of the initial sublattice magnetisation
$m_s(0)$. The lattice size was $349\times 350$ and the results were averaged over
20,000 to 50,000 realisations of the system. \\

\setlength{\unitlength}{0.240900pt}
\ifx\plotpoint\undefined\newsavebox{\plotpoint}\fi
\sbox{\plotpoint}{\rule[-0.200pt]{0.400pt}{0.400pt}}%
%
 \\
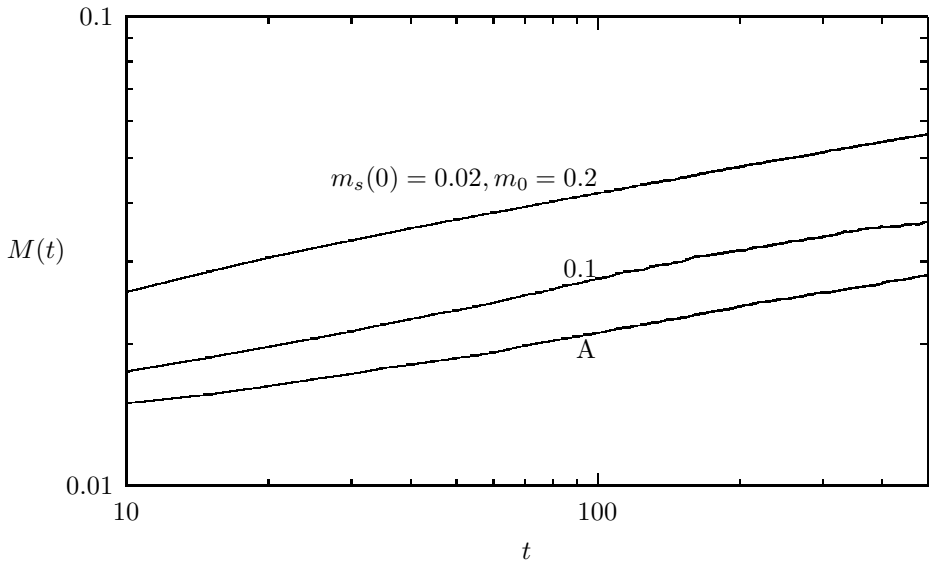
Fig. 2 : Early-time effect at critical temperature in two dimensions for a 
fixed value of the initial sublattice magnetisation $m_s(0)$ and two values of
the conserved magnetisation $m_0$. The line marked A is for Model A, i.e. 
$m_0=0$. The lattice size was $349\times 350$ and the results were averaged over
20,000 to 50,000 realisations of the system. \\

\setlength{\unitlength}{0.240900pt}
\ifx\plotpoint\undefined\newsavebox{\plotpoint}\fi
\sbox{\plotpoint}{\rule[-0.200pt]{0.400pt}{0.400pt}}%
%
 \\
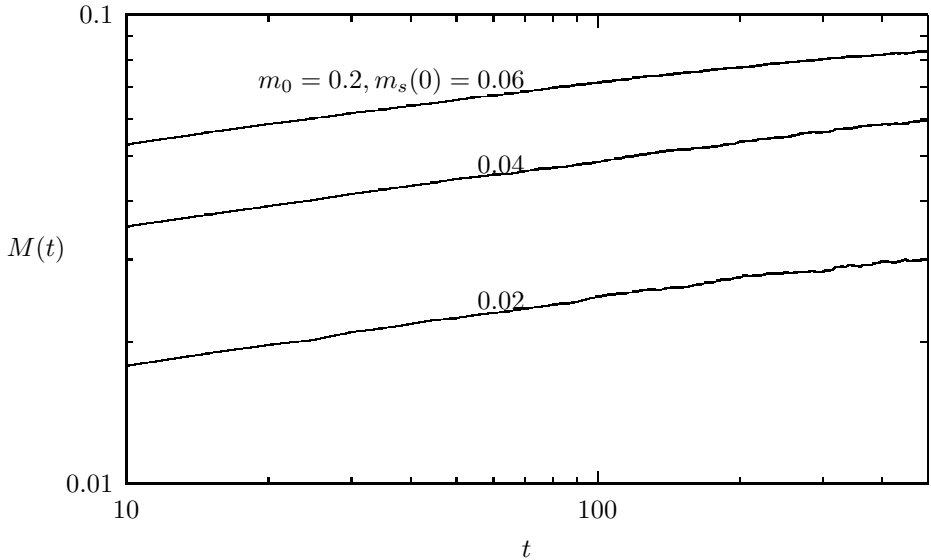
Fig. 3 : Early-time effect at critical temperature in three dimensions for a
fixed value of the conserved
magnetisation $m_0$, and various values of the initial sublattice magnetisation
$m_s(0)$. The lattice size was $65\times65\times 66$ and the results were 
averaged over 5000 realisations of the system. \\

\setlength{\unitlength}{0.240900pt}
\ifx\plotpoint\undefined\newsavebox{\plotpoint}\fi
\sbox{\plotpoint}{\rule[-0.200pt]{0.400pt}{0.400pt}}%
%

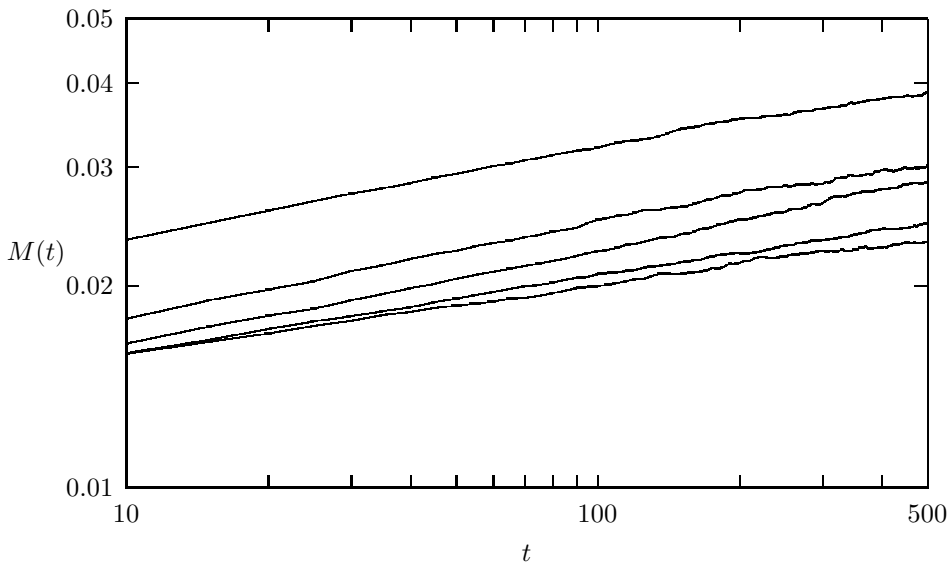
Fig. 4 : Early-time effect at critical temperature in three dimension for a
fixed value of the initial sublattice magnetisation $m_s(0)$ and several 
values of the conserved magnetisation $m_0$. From top to bottom, the 
lines correspond to $m_0=$0.4, 0.2, 0.12, 0.08, 0 (Model A); the results
were averaged over 5000, 5000, 50000, 50000, 13000 realisations respectively.
The lattice size was $65\times65\times 66$ in all the cases. 

\end{document}